# Generation of intermediately-long sea waves by weakly sheared winds

V.M. Chernyavski[1], Y. M. Shtemler[2], E. Golbraikh[3], M. Mond[2]
[1]Institute of Mechanics, Moscow State University, P.O.B. B192, Moscow 119899, Russia,
[2]Department of Mechanical Engineering, [3]Department of Physics,
Ben-Gurion University of the Negev, P.O.B. 653, Beer-Sheva 84105, Israel

The present study concerns the numerical modeling of sea-wave instability under the effect of logarithmic-wind profile in hurricane conditions. The central point of the study is the calculation of the wave growth rate, which is proportional to the fractional input energy from the weakly-sheared (logarithmic) wind to the wave exponentially varying with time. It is shown for hurricane conditions that the Miles-type stability model based on the Charnock's formula with the standard constant coefficient underestimates the growth rate ~5 to 50 times as compared with the model employing the roughness adopted from experimental data for hurricane winds. The drag reduction with wind speed at hurricane conditions coupled with the similar behavior of the dimensionless gravity acceleration, leads to the minimum in the maximal growth rate and the maximum in the most unstable wavelength.

**1. Introduction.** The present study is motivated by recent experimental findings of drag reduction at hurricane conditions.[1-3] It points out the significance of actual experimental parameters of the wind-speed profile for the modeling of the sea-surface instability instead of commonly employed Charnock's relation[4] with the constant proportionality coefficient between roughness and friction-velocity squared.

As speculated in Ref. 1, the foam coverage increase due to wave breaking forms a slip surface at the air-ocean interface that leads to a reduction of the ocean drag at hurricane wind speeds. In Ref. 5, the system has been modelled by a three-fluid system of the foam layer sandwiched between the atmosphere and the ocean, by distributing the foam spots homogeneously over the ocean surface. In the present study, the problem is reduced to modeling of the wind waves over the foam-free portion of the ocean surface, under the effect of the log-wind profile averaged over alternating foam-free and foam-covered portions of the ocean surface.

**2. Physical model.** In general, two distinct prototypical velocity profiles of longitudinal winds are distinguished – weakly- and highly-sheared winds. They are described by unbounded (log-type) and bounded (tanh-type) functions of the distance from the interface:[6,7]

$$\frac{U_a}{U_*} = \log(\frac{y}{L_a}), \qquad y \geq L_a, \qquad (1)$$

$$\frac{U_a}{U_*} = \tanh(\frac{y}{L_a}), \qquad y \geq 0, \qquad (2)$$

where the characteristic wind, $U_*$, and the roughness length, $L_a$, are determined empirically. The hyperbolic tangent profile is frequently used in geophysics and astrophysics for the modeling of sea-waves under shear winds,[7] while the logarithmic profile has been confirmed by recent measurements.[1]

The standard wind-wave stability problem is made dimensionless using water density, $\rho_* = \rho_w$, the characteristic wind speed, $U_*$, additionally specified below, which is a coefficient in (1) or (2), and the gravitational length $L_* = L_g = U_*^2/g$ ($g$ is the gravity acceleration). The resulting dimensionless system contains three dimensionless parameters: the air-water density ratio, the ratios of roughness and capillary length to the gravitational ones:

$$\varepsilon^2 = \frac{\rho_a}{\rho_w}, \quad G = \frac{L_a}{L_g} \equiv \frac{gL_a}{U_*^2}, \quad \Sigma = \frac{L_c}{L_g}. \qquad (3)$$

Here $\varepsilon^2 \approx 10^{-3}$ for the air-water system; $L_c = \sqrt{\sigma/g\rho_w}$ is the capillary length; $\sigma$ is the surface tension; the dimensionless roughness has the physical meaning of the dimensionless gravity acceleration inversely proportional to the Froude number $G=1/Fr$.

Although the main attention will be on the log-like winds, qualitative distinctions in both types of the wind profile will be briefly discussed now. The classic Miles' solution for the sea-wave generation by logarithmic winds was obtained by asymptotic expansions in small ε.[9] As demonstrated recently,[10] Miles' resonant mechanism predicts well the shortwave instability with wave length $\lambda \sim L_g \sim L_c \sim L_a$ ($G \sim 1, \Sigma \sim 1$), generated by weak winds approximated by a piece-wise linear function. For bounded-wind profile (2), the shortwave Miles-type instability induced by weak winds is changed at strong winds ($G \ll 1, \Sigma \ll 1$) by a generalized Kelvin-Helmholtz instability of the intermediately long waves, $\lambda \sim L_g \gg L_c, L_a$. Indeed, at $G \ll 1$ a transition to the characteristic scale $L_* = L_g$ reduces $\tanh(y/L_a)$ in (2) to the Heaviside function uniformly in $G$, and the influence of the roughness is negligibly small. This differs from the strong logarithmic wind (1), which is represented as a sum of $\log(y/L_g)$ and $\ln G$, when both terms should be accounted simultaneously, since $\ln G \sim 1$ for typical values of $G = L_a / L_g \ll 1$ under hurricane conditions. Thus, the wave instability excited by a log-like wind will be described by the Miles-type mode induced as for weak winds by the wind shear effects.

In the present work, intermediately long gravity waves ($\lambda \sim L_g$) are studied for log-type winds (1) under hurricane conditions. In that case the characteristic wind speed $U_* = U_f / \kappa$ may be expressed through roughness for any given neutral stability 10-m wind speed, $U_{10}$:

$$U_f / U_{10} = \kappa / \log(L_{10}/L_a), \quad L_{10} = 10m, \quad (4)$$

where $U_f$ is the friction velocity, $\kappa = 0.4$ is von Karman's constant. Different physically significant characteristics of the logarithmic wind may be expressed to one another for any fixed value of $U_{10}$. For instance, the drag coefficient $C_d = U_f^2 / U_{10}^2$ and the effective gravity acceleration $G = L_a / L_g$ are parametrically related:

$$C_d = \kappa^2 / \log^2 \frac{L_{10}}{L_a}, \quad G = \frac{gL_a}{U_{10}^2} \log^2 \frac{L_{10}}{L_a}. \quad (5)$$

From the first of relations (5) follows that if one of the values $L_a / L_{10}$ and $C_d$ has a maximum, another one should reach its maximum at the same value of $U_{10}$. Since Charnock' coefficient is equal, up to a constant factor, to dimensionless roughness, it may be obtained by comparison of the generalized Charnock' formula[4] with Eq. (3) for $G$ that depends on wind speed:

$$L_a = \alpha_{ch} U_f^2 / g \equiv GU_*^2 / g, \quad G = \alpha_{ch} \kappa^2, \quad (6)$$

where the constant coefficient $\alpha_{ch}$ is commonly adopted from experimental data for low-wind speeds and extrapolated to high wind speeds. Note a large dispersion in the estimations of Charnock's constant coefficient for low-wind speeds, from $\alpha_{ch} \approx 0.01$ taken in Ref. 8 to $\alpha_{ch} \approx 0.015$–0.035 recommended in Ref. 1. Conventional Charnock's formula (6) with $\alpha_{ch} = const$ means the constant effective gravity acceleration $G = \alpha_{ch} \kappa^2$. Below the mean value $\alpha_{ch} = 0.025$ is adopted for estimations. Alternatively, in the present study, $\alpha_{ch} = G / \kappa^2 = gL_a / U_f^2$ in (6) is evaluated with the help of (4) and (5) using actual data for $U_f$ under hurricane conditions, at which $G$ changes with wind speed.[1-3] In the latter case the obtained effective value of wind-speed dependent Charnock's coefficient is almost an order less than the constant values conventionally adopted for weak winds (Table 1).

At hurricane conditions, the gravity length is a natural characteristic scale of the problem and the surface tension effect is negligibly small ($\Sigma \approx 0$), while the density ratio for the air-water system is fixed. Then the effective gravity acceleration, $G = \alpha_{ch} \kappa^2$, or, equivalently, Charnock's coefficient, $\alpha_{ch}$, is the single parameter that governs the system stability, which, in general, depends on the wind speed.

In two dimensions, the governing dimensional equations for linearized

inviscid shear flows with prescribed mean velocities $U_n(y)$ are given by

$$\frac{\partial u_n}{\partial t}+U_n\frac{\partial u_n}{\partial x}+\frac{\partial U_n}{\partial y}v_n=-\frac{1}{\rho_n}\frac{\partial P_n}{\partial x}, \quad (7)$$

$$\frac{\partial v_n}{\partial t}+U_n\frac{\partial v_n}{\partial x}=-\frac{1}{\rho_n}\frac{\partial P_n}{\partial y}-g, \quad (8)$$

$$\frac{\partial u_n}{\partial x}+\frac{\partial v_n}{\partial y}=0, \quad (9)$$

where $n$ is $a$, $w$ for air and water, respectively; $u_n$ and $v_n$ are the horizontal and vertical components of the disturbed velocity, $P_n$ is the pressure, $\rho_n$ is the density. It is assumed here that unperturbed water velocity equals zero $U_w \equiv 0$. Furthermore, solutions for both air and water obey the linearized dynamic and kinematic boundary conditions at the air-water interface, $y = L_a + \zeta$:

$$\Delta P(x,y,t) - \Delta \rho\, g\, \zeta(x,t) = 0, \quad (10)$$

$$\frac{\partial \zeta(x,t)}{\partial t} - v_n(x,y,t) = 0, \quad (11)$$

where $\Delta P = P_a - P_w$, $\Delta \rho = \rho_a - \rho_w$.

Assuming the hydrodynamic motion periodic in time and space yields
$v_n(x, y, t) = W_n(y) E$, $u_n(x,y,t)=(i/\alpha)W_n'(y)E$,
$P_n(x,y,t) = Q_n(y) E$, $\zeta(x,t) = Z E$. (12)
where $E(x,t)=\exp[i(\alpha x - \omega t)]$; $\alpha = 2\pi/\lambda$ is the wavenumber, $\omega = \omega_r + i\omega_i = \alpha C$ and $C = C_r + i C_i$ are the complex frequency and phase velocity. Making use of (12) and eliminating pressure from the problem, (10)-(11) result in Rayleigh's equation:

$$[U_n(y)-C](\frac{d^2W_n}{dy^2}-\alpha^2 W_n)-\frac{d^2U_n}{dy^2}W_n=0. \quad (13)$$

The solution of the Rayleigh equation obeys the corresponding boundary conditions following from (10)-(11). The present calculations are carried out for $\omega_i = \alpha C_i > 0$, outside the critical height, $y_{cr}$, where $C=U_a(y_{cr}/L_a)$, the Rayleigh equation is singular and the Miles's resonance condition is satisfied.

**3. Results of modeling.** A numerical solution of the stability problem adopted here can be easily found for any wind profile.[7] In particular, for the log-type profile, the influence of the roughness is essential even for small $G$ typical for hurricane conditions. The dimensionless (denoted by upper bars) growth rates, $\bar{\omega}_i$, vs. wavelength, $\bar{\lambda}$, calculated for log-profile of the wind, are depicted in Fig. 1 for four hurricane wind speeds. For comparison, in Fig. 2 similar curves are presented for the constant Charnock coefficient in relation (6). Since in the latter case $G$ is independent of $U_{10}$, the single curve for $\bar{\omega}_i$ vs. $\bar{\lambda}$ is depicted in Fig. 2.

Input data and summarized results of calculations are based either on the hurricane data adopted from Ref. 1 (upper values in Table 1), or on Charnock's relation with $\alpha_{ch} = 0.025$ (the lower values in parentheses in Table 1).

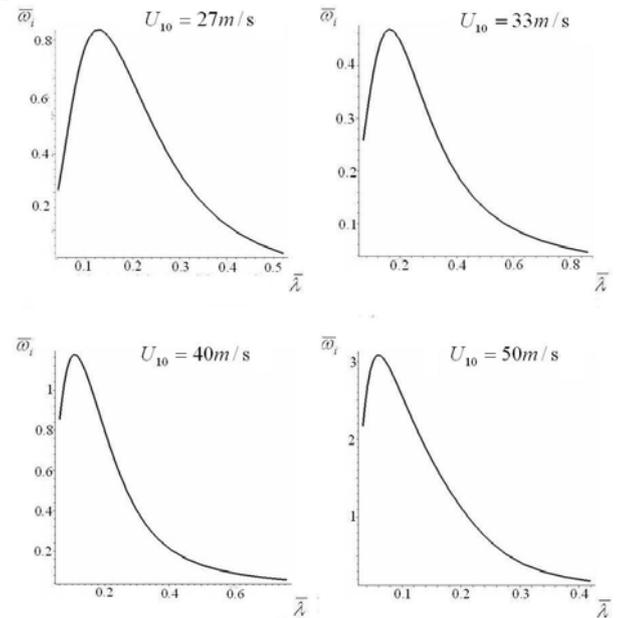

Figure 1. Dimensionless growth rate vs. wavelength ($\bar{\lambda} = \lambda/L_g \lesssim 1$) for four values of $U_{10}$, and the wind-profile parameters adopted from Ref. 1, $\Sigma \approx 0$.

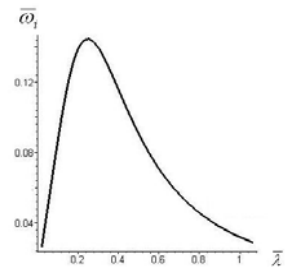

Figure 2. Same as in Fig. 1 for the wind-profile parameters evaluated by Charnock's formula (6) with $\alpha_{ch} = 0.025$, independently of $U_{10}$, $\Sigma \approx 0$.

Some typical dimensionless characteristics of the unperturbed wind and unstable waves are presented in Figs. 3 and 4, respectively (see also Table 1).

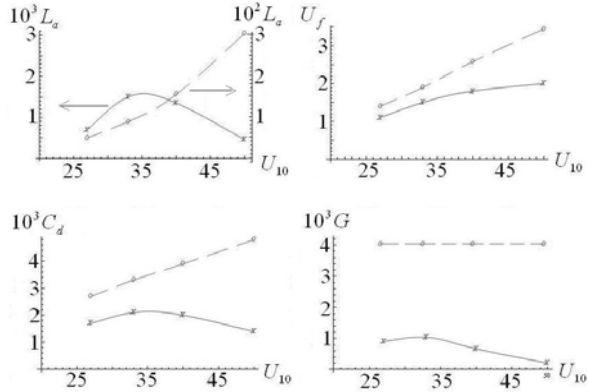

Figure 3. Roughness, $L_a$ [m], friction velocity, $U_f$ [m/s], drag coefficient, $C_d$ and dimensionless gravity acceleration $G = \alpha_{ch}\kappa^2$ vs $U_{10}$ [m/s] for the hurricane wind data[1] at $U_{10}$ =27, 33, 40, 50 [m/s] (solid lines), and for Charnock's formula (6) with $\alpha_{ch} = 0.025$ (dashed lines).

As distinct from the model with constant Charnock's coefficient, note the non-monotonic behavior of dimensionless gravity acceleration $G$, which reflects the non-monotonicity of the drag (roughness) in hurricane conditions (Fig. 3). At relatively weak winds ($U_{10}$<33m/s), the dimensionless gravity acceleration $G$ increases with wind speed $U_{10}$.

Since the gravity acceleration stabilizes the interaction of the atmosphere and the ocean, the growth rate will drop with $G$ rising along with wind speed. With further increase of the wind speed, $G$ falls, and the growth rate rises. This is coupled with the non-monotonic behavior of the dimensionless maximal growth rates and the most unstable wavelengths calculated on the basis of experimental data for hurricane winds[1] (Table 1, Fig. 4). The dimensional phase velocity at small density ratio approximately equals its value at deep water $C \approx C_0 = \sqrt{g\lambda/2\pi}$ (e.g., Eq. (3.7) in Ref. 11). This provides a non-monotonic behavior of phase velocity and frequency with the wind speed, $U_{10}$. Thus, at strong winds, Miles' mode of the linear instability supports the growth of intermediately long waves with wavelengths of the order of $L_g$: $\lambda \le \lambda_{max} \sim 0.2 m \lesssim L_g \sim 1 \div 2\, m$.

For comparison, note that the stability characteristics calculated using roughness obtained via Charnock's formula with $\alpha_{ch} = const$ are constant in wind speed (Table 1, Fig. 4). Furthermore, the maximal growth rates based on the constant-coefficient Charnock's formula are less than those obtained for real hurricane conditions by the factor $K = \omega_i^{max} / (\omega_i^{max})$, which varies from ~5 to ~50 for $U_{10}$ within the range of 27 to 50 m/s. The numerator of $K$ is calculated for the data adopted from Ref. 1, while the denominator is obtained based on Charnock's formula with $\alpha_{ch} = 0.025$.

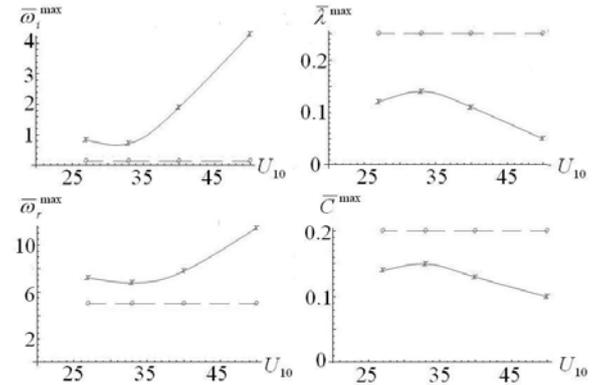

Figure 4. Maximal growth rate, $\overline{\omega}_i^{max}$, and the corresponding wavelength, $\overline{\lambda}^{max}$, frequency, $\overline{\omega}_r^{max}$ and phase velocity, $\overline{C}^{max}$, vs $U_{10}$ [m/s] for the hurricane wind data[1] at $U_{10}$ =27, 33; 40, 50 [m/s] (solid lines) and for Charnock's formula (6) with $\alpha_{ch} = 0.025$ (dashed lines).

**4. Discussion and Summary.** In Ref. 1, the logarithmic wind speed measured at a distance from the ocean surface has been extrapolated to its zero value at the roughness height near the ocean surface ($y = L_a$) in order to evaluate the effective roughness in storm and hurricane conditions. This procedure completely determines the log-wind profile (1) with the help of relations (4)-(6) between the roughness, friction velocity and drag coefficient. Since the drag coefficient estimated in Ref. 1 evidently has an average meaning over alternating foam-free and foam-covered portions of the ocean surface, this procedure also provides an averaged log-wind profile. Since at $U_{10}$<50$m/s$ the ocean surface is incompletely covered by the foam (Fig. 4 in Ref. 1, see also Ref. 12), the averaged log-wind profile is applied here to the modeling of the wind waves over the foam-free portion of the ocean surface.

Since the data in Fig. 3 of Ref. 1 have some freedom of choice in regard to the roughness value, in the present study the experimental data for the friction velocity as the primary variable have been adopted, which are characterized by a much smaller dispersion about the mean value. After that the roughness values are calculated by substituting the mean values of the friction velocity into the basic relation (4).

Since the gravitational length is a single dimensional-length parameter in the stability problem, it can be concluded on the dimensional grounds that the gravity length ($L_g = U^2_*/g$) is a natural scale for the dimensional wavelength $\lambda$. Hence, at storm and hurricane conditions under consideration ($U_* \sim$3-5$m/s$, Table 1), Miles' mode is responsible for the growth of the waves, mainly with the lengths $\lambda$ of the order of or less than $L_g \sim 1 m$ (see Figs. 1-2). Since waves in tropical cyclones, in which the experiments of Powell et al.[1] were performed, may reach the length of a few hundred meters, let us estimate the growth rates of long length Miles' modes, which are characterized by much smaller growth rates, $\bar{\omega}_i \ll 1$ (see Fig. 5). This means that long waves will travel almost with no amplification for a long time $1 \ll \bar{t} \ll 1/\bar{\omega}_i$.

The present study provides a physically transparent relation between the concept of the drag reduction[1-2] and the results of the stability analysis. Indeed, since the drag and roughness reach their maxima at the same value of the wind speed, the non-monotony of the drag coefficient in wind speed leads to non-monotonic behavior of the dimensionless gravity acceleration. Hence, because of the evident stabilizing property of the gravity acceleration in the stably-stratified system of the atmosphere and the ocean, this evidently provides the non-monotony of the growth rate coefficient and the corresponding, phase velocity and frequency.

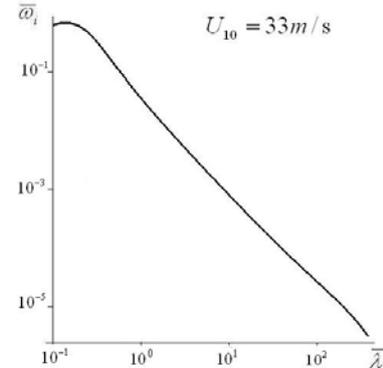

Figure 5. Dimensionless growth rate vs. wavelength (from $\bar{\lambda} = \lambda/L_g \lesssim 1$ up to $\bar{\lambda} = \lambda/L_g \gg 1$) for wind-profile parameters adopted from Ref. 1, $\Sigma$=0.

It is demonstrated that the effective gravity acceleration, $G$, is proportional to Charnock's coefficient, $\alpha_{ch}$, with the proportionality factor equal to the square of Karman's constant. The growth rates calculated using the measured data for hurricane winds, strongly differ from those based on constant Charnock's coefficient. Although this result can be expected because of a significant difference between the corresponding roughness values,[1] the present modeling demonstrates additionally the non-monotonic behavior of the growth rates of the most unstable waves and others stability characteristics. The account for the proportionality of Charnock's coefficient to the effective gravity acceleration, its role of the single dimensionless parameter determining the surface-wave stability and its dependence on wind speed may be useful for the parametrization of roughness in hurricane prediction models in order to provide their coupling with the surface-wave instabilty models.

**References**
[1]Powell, M.D., Vickery, P.J., and Reinhold T.A., Reduced drag coefficient for high wind speeds in tropical cyclones. *Nature*, 422, 279 (2003).

[2] Donelan, M. A. B. K. Haus, N. Reul, W. J. Plant, M. Stiassne, H. C. Graber, O. B. Brown, and E. S. Saltzman, On the limiting aerodynamic roughness of the ocean in very strong winds. *Geophys. Res. Lett.*, 31, L18306 (2004).

[3] E. Jarosz, D. A. Mitchell, D. W. Wang, W. J. Teague, Bottom-Up Determination of Air-Sea Momentum Exchange Under a Major Tropical Cyclone, *Science*, **315**,1707 (2007).

[4] Charnock, H., Wind stress on a water surface. Q. J. R. Met. Soc., **81**, 639(1955).

[5] Shtemler, Y. M. Golbraikh E., Mond M, Wind-wave stabilization by a foam layer between the atmosphere and the ocean, Dyn. of Atmospheres and Oceans, 50(1), 1 (2010)

[6] Morland L. C. and Saffman P. G., Effect of wind profile on the instability of wind blowing over water, J. Fl. Mech., **252**, 383 (1993).

[7] Alexakis A., Young Y., and Rosner R., Shear instability of fluid interfaces: Stability analysis, Phys. Rev. E **65**, 026313 (2002).

[8] Smith, S. D., Wind Stress and Heat Flux over the Ocean in Gale Force Winds, J. Phys. Oceanog. 10, 709 (1980).

[9] Miles J.W., On the wave generation of surface waves by shear flows, J. Fl. Mech., **3**, 185 (2004)

[10] Shtemler, Y. M, Mond M., Chernyavski V., Golbraikh E., Nisim Y., An asymptotic model for the Kelvin-Helmholtz and Miles mechanisms of water wave generation by wind, Phys. Fl., 20(9), 094106-11 (2008).

[11] Bedji S., and Nadaoka K., Solution of Rayleigh' instability equation for arbitrary wind profiles, J. Fl. Mech., **500**, 65 (2004).

[12] S. El-Nimri, W. Jones, E. Uhlhorn, C. Ruf, J. Johnson, and P. Black, An Improved C-Band Ocean Surface Emissivity Model at Hurricane-Force Wind Speeds Over a Wide Range of Earth Incidence Angles, IEEE Geosci. Remote Sens. Lett, 7, 4, 641 (2010)

| $U_{10}$ [m/s] | 27 | 33 | 40 | 50 | $U_{10}$ [m/s] | 27 | 33 | 40 | 50 |
|---|---|---|---|---|---|---|---|---|---|
| $U_f$ [m/s] | 1.1 (1.4) | 1.5 (1.9) | 1.8 (2.5) | 2.00 (3.45) | $\overline{\omega}_i^{max}$ | 0.83 (0.14) | 0.72 (0.14) | 1.9 (0.14) | 4.3 (0.14) |
| $10^3 L_a$ [m] | 0.7 (5.2) | 1.5 (9.0) | 1.35 (15.6) | 0.46 (30.4) | $\overline{\lambda}^{max}$ | 0.12 (0.25) | 0.14 (0.25) | 0.11 (0.25) | 0.05 (0.25) |
| $U_* = U_f/\kappa$ [m/s] | 2.75 (3.5) | 3.75 (4.75) | 4.5 (6.25) | 5.0 (8.6) | $\overline{\omega}_r^{max}$ | 7.2 (5.0) | 6.8 (5.0) | 7.8 (5.0) | 11.5 (5.0) |
| $L_* = L_g$ [m] | 0.77 (1.25) | 1.43 (2.3) | 2.1 (3.99) | 2.2 (7.55) | $\overline{C}_0$ | 0.14 (0.2) | 0.15 (0.2) | 0.13 (0.2) | 0.1 (0.2) |
| $t_* = L_*/U_*$ [s] | 0.28 (0.36) | 0.38 (0.48) | 0.47 (0.63) | 0.48 (0.89) | $\lambda^{max} = \overline{\lambda}^{max} L_*$ [m] | 0.1 (0.3) | 0.2 (0.57) | 0.23 (1.0) | 0.13 (1.9) |
| $10^3 G$ | 0.91 (4.0) | 1.04 (4.0) | 0.67 (4.0) | 0.21 (4.0) | $\omega_r^{max} = \overline{\omega}_r^{max}/t_*$ [1/s] | 25.6 (15.0) | 17.9 (10.4) | 16.4 (8.7) | 23.9 (6.2) |
| $10^3 C_d$ | 1.7 (2.8) | 2.1 (3.3) | 2.0 (3.8) | 1.4 (4.8) | $\omega_i^{max} = \overline{\omega}_i^{max}/t_*$ [1/s] | 3.0 (0.4) | 1.9 (0.3) | 4.0 (0.22) | 8.9 (0.16) |
| $10^2 \alpha_{ch}$ | 0.56 (2.5) | 0.65 (2.5) | 0.42 (2.5) | 0.13 (2.5) | $C_0$ [m/s] | 0.39 (0.7) | 0.56 (0.95) | 0.6 (1.27) | 0.45 (1.7) |
| | | | | | $K = \dfrac{\omega_i^{max}}{(\omega_i^{max})}$ | 7.5 | 6.3 | 18.8 | 55.6 |

Table 1. Equilibrium and stability characteristics at hurricane winds for roughness adopted from Ref. 1 (upper values) and (lower values in brackets) evaluated by Charnock's formula with $\alpha_{ch} \approx 0.025$. Stars and bars denote the characteristic scales and the dimensionless variables